\documentclass[review,12pt]{elsarticle}
\usepackage{lineno,hyperref}
\modulolinenumbers[5]

\journal{Journal of \LaTeX\ Templates}

\usepackage{amsmath}
\usepackage{graphicx}

\title{Acoustomagnetoelectric Effect in Graphene Nanoribbon in the Presence of 
External Electric and Magnetic Fields }

\author[rvt]{Kwadwo A. Dompreh\corref{cor1}}
\author[rvt]{Samuel Y. Mensah}
\address[rvt]{Department of Physics, College of Agriculture and Natural Sciences, U.C.C, Ghana.}
\cortext[cor1]{Corresponding author\\ Email: kwadwo.dompreh@ucc.edu.gh}
\ead[url]{kwadwo.dompreh@ucc.edu.gh}
\author[rvt]{Sulemana S. Abukari}
\author[rvt]{Raymond Edziah}
\author[focal]{Natalia G. Mensah}
\address[focal]{Department of Mathematics, College of Agriculture and Natural Sciences, U.C.C, Ghana}
\author[els]{Harrison A. Quaye}
\address[els]{Department of Computer Science, College of Agriculture and Natural Sciences, U.C.C, Ghana}

\date{}

\begin{document}
\begin{abstract}
\noindent Acoustomagnetoelectric Effect (AME) in Graphene Nanoribbon (GNR) in the 
presence of an external electric and magnetic fields  was studied using the Boltzmann kinetic 
equation. On open circuit, the Surface Acoustomagnetoelectric field ($\vec{E}_{SAME}$) 
in GNR  was obtained in the region $ql >> 1$,  for  energy dispersion $\varepsilon(p)$ 
near the Fermi level. The dependence of $\vec{E}_{SAME}$ on  the magnetic field 
strength ($\eta$), the sub-band index ($p_i$), and  the width ($N$) of GNR  were analysed numerically.
For $\vec{E}_{SAME}$ versus $\eta$, a non-linear graph was obtained. From the graph,
at low magnetic field strength ($\eta < 0.62$), the obtained graph qualitatively agreed with that experimentally observed 
in graphite. However, at high magnetic field strength ($\eta > 0.62$),
the $\vec{E}_{SAME}$ falls rapidly to a minimum value. We observed that in GNR, the maximum
$\vec{E}_{SAME}$ was obtained at magnetic field $H = 3.2 Am^{-1}$. 
The graphs obtained were modulated by varying the sub-band index $p_i$ 
with an inversion observed when $p_i = 6$. The dependence of $\vec{E}_{SAME}$ on the width $N$ for various $p_i$ 
was also studied where, $\vec{E}_{SAME}$ decreases for increase in $p_i$. To enhanced the understanding of $\vec{E}_{SAME}$ on the $N$ and $\eta$, a $3$D graph  was plotted.
This study is relevant for investigating the properties of GNR. 
\end{abstract}

\maketitle
\section*{Introduction}
The study of Acoustomagnetoelectric Effect (AME) in Semiconductors and its related materials have 
generated lot of interest recently . AME in materials such as Superlattices \cite{1, 2, 3}, Quantum Wires \cite{4}, 
Carbon Nanotubes \cite{5} deals with appearance of a d.c electric field in the Hall direction when the 
sample is on open circuit. Studies have shown that the propagation of acoustic waves causes the 
transfer of energy and momentum to the conducting electrons \cite{3}. When the build up of the acoustic 
flux  exceeds the velocity of sound it causes the formation and propagation 
of Acoustoelectric field \cite{6,7}.  Other effects such as Acoustoelectric Effect (AE) \cite{1, 2,8}, Acoustothermal 
Effect \cite{9}, and Acoustoconcentration Effect  can occur. The AE was predicted by Grinberg and Kramer 
\cite{10} for bipolar semiconductors and experimentally observed in Bismuth by Yamada \cite{11}. 
 By applying the sound flux ($\vec{W}$), electric current ($\vec{j}$), and magnetic fields ($\vec{H}$)  perpendicularly to 
the sample, it is interesting to note that, with the sample opened in direction perpendicular to the Hall direction, can leads to a non-zero 
Acoustomagnetoelectric Effect AME \cite{12}. Mensah et. al \cite{1} studied these effect in Superlattice in the 
hypersound regime, Bau et. al.~\cite{13}  studied the AME of cylindrical quantum wires. Also, 
AME effect in mono-polar semiconductor for both weak and quantizing field were studied \cite{14}.
Experimentally, AME has been observed in n-InSb \cite{15}, and in graphite \cite{16} for $ql << 1$.  
In this paper, AME in graphene nanoribbon is studied. There are differences between graphene and graphite. 

Graphene\cite{17} is a single layer of carbon atoms with zero band-gap.  Within the low energy 
range ($\varepsilon < 0.5 eV$), carriers in graphenes are massless relativistic particles with effective speed of  
$V_F\approx 10^6 ms^{-1}$ ($V_F$ being the Fermi velocity).  One of the major limitations of  graphene sheet is  lack of  
band gap in its energy spectrum~\cite{18}. To overcome this, stripes of Graphene called 
Graphene Nanoribbons (GNRs) whose characteristics are dominated by the nature of their edges (the armchair  (AGNRs) and 
Zigzag (ZGNRs)) with well-defined width are proposed \cite{18}.  By patterning graphene into narrow ribbons creates an energy gap 
where GNR behaves like semiconductor \cite{19,20,21}.  However, graphite (bunch of graphene) have planar structures with a semimetallic 
behaviour having a band overlap of about $4.1MeV$. Its thermal, acoustic and electronic properties are highly anisotropic, which 
means that phonons travel much easily along the planes than they do through the planes~\cite{23}. Graphene therefore have 
a very high electron mobility thus offers a much better level of electronic  conduction.
In this paper,  the Boltzmann kinetic equation is used to study the SAME in GNR.  This is achieved by applying sound 
flux ($\vec{W}$) to the GNR sample in the presence of electric field ($\vec{E}$) and magnetic fields ($\vec{H}$). With 
the sample open ($j = 0$), give the $E_{SAME}$ in GNR.
This paper is organised as follows: In section 2, the theory of SAME in GNRs is outlined. In section 3, the numerical calculations  
are presented; and  while section 4 deals with the conclusion. 

\section*{Theory}
The configuration for suface Acoustomagnetoelectric field  in GNR will be considered with the acoustic phonon $\vec{W}$,
the magnetic field $\vec{H}$ and the measured $E_{SAME}$ lying in the same plane. Based on the method developed in \cite{22}, 
the partial current density generated in the sample is solved from the Boltzmann transport equation given as 
\begin{eqnarray}
 -\left(e\vec{E}\frac{\partial f_{\vec{p}}}{\partial{\vec{p}}} + \Omega[\vec{p},\vec{H}],\frac{\partial f_{\vec{p}}}{\partial {\vec{p}}}\right)=
 -\frac{f_{\vec{p}} - f_0(\varepsilon_{\vec{p}})}{\tau(\varepsilon_{\vec{p}})}  +\nonumber\\
  \frac{\pi {\Delta}^2 \vec{W}}{\rho V_s^3} \left\{ {[f_{{\vec{p}}+{\vec{q}}}} - f_{\vec{p}}]\delta(\varepsilon_{\vec{p}+\vec{q}} -\varepsilon_{\vec{p}}-\hbar\omega_{\vec{q}})+ {[f_{{\vec{p}}-{\vec{q}}}}
 - f_{\vec{p}}]\delta(\varepsilon_{\vec{p}-\vec{q}} -\varepsilon_{\vec{p}}+\hbar \omega_{\vec{q}})\right\}\label{EQ_1} 
\end{eqnarray}
where $ql >> 1$ is utilised. Here, $f_0{(\varepsilon(\vec{p}))}$ is the equilibrium distribution function,  $\vec{E}$ is the 
constant electric field, $\omega_{\vec{q}}$ is the fequency of the acoustic wave, $\vec{W}$ is the 
density of the acoustic flux, and $\vec{p}$ the characteristic quasi-momentum of the electron. $\rho$ is the density of the 
sample, $\Delta$ is the constant of deformation potential, $e$ the electronic charge, and $V_s$ is the speed of sound. 
The relaxation time on energy is $\tau(\varepsilon_{\vec{p}})$ and the cyclotron frequency, $\Omega = {\mu H}/{\hbar c}$
($H$ is the magnetic field, $\mu$ is the electron mobility and $c$ is the speed of light in vacuum). The energy dispersion relation $\varepsilon(\vec{p})$
for GNRs band near the Fermi point is expressed  as \cite{18,24}
\begin{equation}
 \varepsilon(\vec{p}) = \frac{E_g}{2} \sqrt{[(1+\frac{\vec{p}^2}{{\hbar}^2{\beta}^2})]} \label{EQ_2}
\end{equation}
where the energy gap $E_g = 3ta_{c-c}\beta$ with $\beta$ being the quantized wave vector given as $\beta = \frac{2\pi}{a\sqrt{3}}[\frac{p_i}{N+1}-\frac{2}{3}]$, 
where  $p_i$ is the sub-band index and $N$ is the width of the GNR. $t =2.7$eV is the nearest neighbour 
Carbon-Carbon C-C tight binding overlap energy and $a_{c-c} = 1.42\dot{A}$ is the (C-C) bond length.

Multiplying the Eqn.(\ref{EQ_1}) by $\vec{p}\delta(\varepsilon - \varepsilon_{\vec{p}})$ and summing over
$\vec{p}$ gives the kinetic equation as 
\begin{equation}
 \frac{{\vec{R}(\varepsilon)}}{\tau(\varepsilon)} + \Omega\left[{\vec{h}},{\vec{R}(\varepsilon)}\right] = {\vec{\Lambda}(\varepsilon)}+{\vec{S}(\varepsilon)}  \label{EQ_3}
 \end{equation}
where ${\vec{R}(\varepsilon)}$ is the partial current density given as 
\begin{equation}
{\vec{R}}(\varepsilon) \equiv e\sum_{\vec{p}} \vec{p} f_{\vec{p}} \delta(\varepsilon -\varepsilon_{\vec{p}}) \label{EQ_4}
\end{equation}
with $\vec{\Lambda}(\varepsilon)$ and ${\vec{S}}{(\varepsilon)}$ given as
\begin{equation}
{\vec{\Lambda}}(\varepsilon) =-e\sum_{\vec{p}}\left({\vec{E}},\frac{\partial f_{\vec{p}}}{\partial{\vec{p}}}\right){\vec{p}}\delta(\varepsilon -\varepsilon_{\vec{p}}) \label{EQ_5}
\end{equation}
\begin{eqnarray}
{\vec{S}}{(\varepsilon)} = \frac{\pi {\Delta}^2 \vec{W}}{\rho V_s^3}\sum_{\vec{p}}\vec{p}\delta(\varepsilon -\varepsilon_{\vec{p}})\{ {[f_{{\vec{p}}+{\vec{q}}}} - f_{\vec{p}}]\delta(\varepsilon_{\vec{p}+\vec{q}} -\varepsilon_{\vec{p}}-\hbar\omega_{\vec{q}})
 +{[f_{{\vec{p}}-{\vec{q}}}} - f_{\vec{p}}]\nonumber\\
\delta(\varepsilon_{\vec{p}-\vec{q}} -\varepsilon_{\vec{p}}+\hbar\omega_{\vec{q}})\}  \label{EQ_6}
\end{eqnarray}
Considering $f_{\vec{p}} \rightarrow  f_0(\varepsilon_{\vec{p}})$ with ${\vec{p} \rightarrow -{\vec{p}}}$ ,
$f_{\vec{p}} \equiv f_0{(\varepsilon_{\vec{p}})} = f_0{(\varepsilon_{-\vec{p}})} $, Eqn.(\ref{EQ_5}) and Eqn.(\ref{EQ_6}) can be respectively expressed to 
\begin{equation}
\vec{\Lambda}(\varepsilon) =\vec{E}\left(\frac{2{\hbar}^2\beta^2}{{\hbar}\vec{q}}\alpha - \frac{{\hbar}\vec{q}}{2}\right)\frac{\partial {f_0}}{\partial{\varepsilon}} \frac{\Theta \left(1 -{\alpha}^2\right)}{\sqrt{1- {\alpha}^2}}\label{EQ_7}
\end{equation}
\begin{eqnarray}
{\vec{S}}({\varepsilon}) = \frac{2\pi{{\vec{W}}}}{\rho V_s\alpha}\Gamma_0\left(\frac{2\hbar^2\beta^2}{\hbar \vec{q}}\alpha -\frac{\hbar \vec{q} }{2}\right)
 \frac{\Theta \left(1 -{\alpha}^2\right)}{\sqrt{1- {\alpha}^2}}\frac{1}{f_0(\varepsilon)}\frac{\partial f_0}{\partial {\varepsilon}} \label{EQ_8}
\end{eqnarray}
with $\alpha = {\hbar\omega_{\vec{q}}}/{E_g}$,  $\Gamma_0 =({E_g^2 {\Delta}^2 \alpha^2}/{2 V_s^2})f_0(\varepsilon)$ and $\Theta$ is the Heaviside step function given as
\begin{equation*}
\begin{centering}
\Theta(1-\alpha^2)=
\begin{cases}{
1 \quad \text{ if $(1-\alpha^2) > 0$}}\\
{0 \quad \text{if $(1-\alpha^2) < 0$}
}\end{cases}
\end{centering}
\end{equation*}
Substituting Eqn.(\ref{EQ_7}) and Eqn.(\ref{EQ_8}) into Eqn.(\ref{EQ_3}) and solving for $\vec{R}(\varepsilon)$ gives
\begin{eqnarray} 
\vec{R}(\varepsilon) =\{\frac{2\pi}{\rho
V_s\alpha}\Gamma_0\left(\frac{2\hbar^2\beta^2}{\hbar \vec{q}}\alpha -\frac{\hbar \vec{q}}{2}\right) \frac{\Theta \left(1 -{\alpha}^2\right)}{\sqrt{1- {\alpha}^2}}
\frac{1}{f_0(\varepsilon)}\frac{\partial f_0}{\partial {\varepsilon}}\times\nonumber\\
\{\vec{W}\tau(\varepsilon)+{\Omega}[\vec{h},\vec{W}]\tau(\varepsilon)^2 + \Omega^2\vec{h}(\vec{h},\vec{W})\tau(\varepsilon)^3\}+\left(\frac{2\hbar^2\beta^2}{\hbar \vec{q}}
\alpha -\frac{\hbar \vec{q} }{2}\right)\frac{\partial f_0}{\partial {\varepsilon}} \frac{\Theta \left(1 -{\alpha}^2\right)}{\sqrt{1- {\alpha}^2}}\times\nonumber\\
\{\vec{E}\tau(\varepsilon)+\Omega[\vec{h},\vec{E}]\tau(\varepsilon)^2 +  \Omega^2 \tau(\varepsilon)^3 \vec{h}(\vec{h},\vec{E})\}\}\{1+\Omega^2\tau(\varepsilon)^2\}^{-1}\label{EQ_12}
\end{eqnarray}
 The current density~\cite{6} is given as 
\begin{equation}
\vec{j}=-\int_0^{\infty}{\vec{R}(\varepsilon) d\varepsilon} \label{EQ_13}
\end{equation}
With $\Delta = \left(\frac{2\hbar^2\beta^2}{\hbar \vec{q}}\alpha -\frac{\hbar \vec{q}}{2}\right)$, substituting Eqn.(\ref{EQ_12}) into Eqn.(\ref{EQ_13}) yields
\begin{eqnarray} 
\vec{j} =\frac{ \Delta\Gamma_0}{\rho V_s \alpha}\frac{\Theta \left(1 -{\alpha}^2\right)}{\sqrt{1- {\alpha}^2}}\{\langle\langle\frac{\tau(\varepsilon)}{1+\Omega^2\tau(\varepsilon)^2}\rangle\rangle {\vec{W}}+
{\Omega}\langle\langle\frac{\tau(\varepsilon)^2}{1+\Omega^2\tau(\varepsilon)^2}\rangle\rangle[\vec{h},\vec{W}] + \nonumber\\
{\Omega^2}\langle\langle\frac{\tau(\varepsilon)}{1+\Omega^2\tau(\varepsilon)^2}\rangle\rangle {\vec{h}(\vec{h},\vec{W})}\}+
\Delta\frac{\Theta \left(1 -{\alpha}^2\right)}{\sqrt{1- {\alpha}^2}}\{\langle\frac{\tau(\varepsilon)}{1+\Omega^2\tau(\varepsilon)^2}\rangle{\vec{E}}+{\Omega}\langle\frac{\tau(\varepsilon)^2}{1+\Omega^2\tau(\varepsilon)^2}\rangle[\vec{h},\vec{E}]+\nonumber\\
{\Omega^2}\langle\frac{\tau(\varepsilon)^3}{1+\Omega^2\tau(\varepsilon)^2}\rangle \vec{h}(\vec{h},\vec{E})\}\label{EQ_14}
\end{eqnarray}
The Eqn.(\ref{EQ_14}) can further be simplified with the following substitution  $g = {1}/{1+\Omega^2\tau(\varepsilon)^2}$, 
$\gamma_k \equiv\langle g\tau(\varepsilon)^k\rangle$, and $\eta\equiv\langle\langle g\tau(\varepsilon)^k\rangle\rangle$ 
where $k = 1, 2, 3$. This yields
\begin{eqnarray}
\vec{j} =\frac{\Delta\Gamma_0}{\rho V_s\alpha}\frac{\Theta(1-\alpha^2)}{\sqrt{1-\alpha^2}}\left\{\eta_1\vec{W} + \Omega\eta_2[\vec{h},\vec{W}]+\Omega^2\eta_3\vec{h}(\vec{h},\vec{W} )\right\}+ \nonumber\\
 \Delta\frac{\Theta \left(1 -{\alpha}^2\right)}{\sqrt{1- {\alpha}^2}}\left\{\gamma_1\vec{E}+\gamma_2\Omega[\vec{h},\vec{E}]+\Omega^2\gamma_3\vec{h}(\vec{h},\vec{E})\right\}\label{EQ_15}
\end{eqnarray}
With the sample opened ($\vec{j}=0$), and ignoring higher powers of $\Omega$ gives 
\begin{eqnarray}
\gamma_1 \vec{E}_x - \gamma_2\Omega\vec{E}_y&=&-\gamma_1\vec{E}_{\alpha}\\ \label{EQ_16}
\gamma_2\Omega \vec{E}_x + \gamma_2\Omega\vec{E}_y&=&-\gamma_2\Omega\vec{E}_{\alpha}\label{EQ_17}
\end{eqnarray}
where $E_{\alpha}= \frac{\Gamma_0}{\rho S\alpha}$. Making the $\vec{E}_y$ the subject of the equation yields
\begin{equation}
\vec{E}_y = \vec{E}_{\alpha}\Omega\left\{\frac{\eta_1 \gamma_2 - \eta_2 \gamma_1}{\gamma_1^2 + \gamma_2^2\Omega^2}\right\}\label{EQ_18}
\end{equation}
substituting the expressions for $\eta_1, \eta_2, \gamma_1, \gamma_2$ into Eqn.(\ref{EQ_18}), with $\vec{E_y}=\vec{E}_{SAME}$ gives 
\begin{equation}
\vec{E}_{SAME}=\vec{E}_{\alpha}\Omega
\left\{\frac{\langle\frac{\tau(\varepsilon)^2}{1+\Omega^2\tau(\varepsilon)^2}\rangle 
\langle\langle\frac{\tau(\varepsilon)}{1+\Omega^2\tau(\varepsilon)^2}\rangle\rangle-
\langle\langle\frac{\tau(\varepsilon)^2}{1+\Omega^2\tau(\varepsilon)^2}\rangle\rangle
\langle\frac{\tau(\varepsilon)}{1+\Omega^2\tau(\varepsilon)^2}\rangle }
{{\langle\frac{\tau(\varepsilon)}{1+\Omega^2\tau(\varepsilon)^2}\rangle}^2+
{\langle\frac{\tau(\varepsilon)^2}{1+\Omega^2\tau(\varepsilon)^2}\rangle}^2\Omega^2 }\right\}\label{EQ_19}
 \end{equation}
In Eqn($16$), the following averages were used
\begin{displaymath}
\begin{centering}
\begin{aligned}
 \langle....\rangle
 &=-{\int_{0}^{\infty}} (....)\frac{\partial f_0}{\partial{\varepsilon}}d{\varepsilon}\\
\langle\langle....\rangle\rangle
  &=-\frac{2\pi}{f_0({\varepsilon})}{\int_{0}^{\infty}} (....)\frac{\partial f_0}{\partial{\varepsilon}}d{\varepsilon}
\end{aligned}
\end{centering}
\end{displaymath}
Where $f_0 = [ 1 - exp(-\frac{1}{kT}(\varepsilon - \varepsilon_F))]^{-1}$ is the Fermi-Dirac distribution function.  
\section*{Numerical analysis and Discussions}
In solving for Eqn.({\ref{EQ_19}}), the following were assumed: At low temperature $kT << 1$, and $\frac{\partial f_0}{\partial \varepsilon} = 
\frac{-1}{k_{\beta}T}exp(-\frac{\varepsilon-\mu}{k_{\beta} T})$.
The equation for $\vec{E}_{SAME}$  simplifies to
\begin{eqnarray}
\vec{E}_{SAME} =\frac{E_g\vec{W}\hbar\omega_{\vec{q}}\eta}{2\rho V_s^3}
\left\{F_{(-1/2,\eta^2)}F_{(-3/2,\eta^2)}-F_{(0,\eta^2)}F_{(-2,\eta^2)}\right\}\times\nonumber\\
\left\{ \frac{3\sqrt{\pi}}{4}F_{(-1/2,\eta^2)}^2 + \frac{9\pi}{16}\eta^2F_{(0,\eta^2)}^2\right\}^{-2} \label{EQ_20}
\end{eqnarray}
with $F_{m,n} = \int_0^{\infty}{\frac{x^m}{1+ \Omega^2\tau(\varepsilon)^2x^n}}\frac{\partial f_0{(\varepsilon)}}{\partial x}dx$. 
From Eqn.(\ref{EQ_20}), the $\vec{E}_{SAME}$ is a function of the following parameters: magnetic field strength ($\eta = \Omega\tau$); 
$\alpha $; and the energy gap $E_g = 3t a_{c-c}\beta$. The $E_g$ depends on the quantized wave vector 
$\beta$. The parameters used in the numerical calculations are $\tau = 10^{-12}s$, $\omega_q = 10^{10}s^{-1}$, $s = 5*10^3ms^{-1}$, 
$q = 2.23*10^6 cm$. In analysing the Eqn.(\ref{EQ_20}), the condition ($(1- \alpha^2)> 0$) was considered.
Figure 1a, shows the dependence of $\vec{E}_{SAME}$  against the magnetic field strength $\eta$ at various sub-bands for $\eta << 1$. 
Generally, $\vec{E}_{SAME}$  increased to a maximum value for three different values of $p_i$. The results obtained 
(see Figure $1a$) qualitatively agreed with an experrimental graph measured in graphite.  Figure $1b$ is 
the general case when there is no limitation on $\eta$.
It can be seen that, $\vec{E}_{SAME}$ decreased rapidly after the maximum point to a minimum value.  For $p_i = 6$, there
is an inversion of the graph.  Figure 2,  shows the dependence of $\vec{E}_{SAME}$ against the width $N$ with 
different sub-band indices ($p_i$). For further illucidation of the graphs obtained, a $3$D graph of  $\vec{E}_{SAME}$ versus $\eta$ 
at $p_i = 1$ and width at $p_i = 6$  are  presented (see Figure 3 a and b) where Figure $3b$ shows  an 
inversion of Figure 3a. 
\begin{figure}
\includegraphics[width = 7cm]{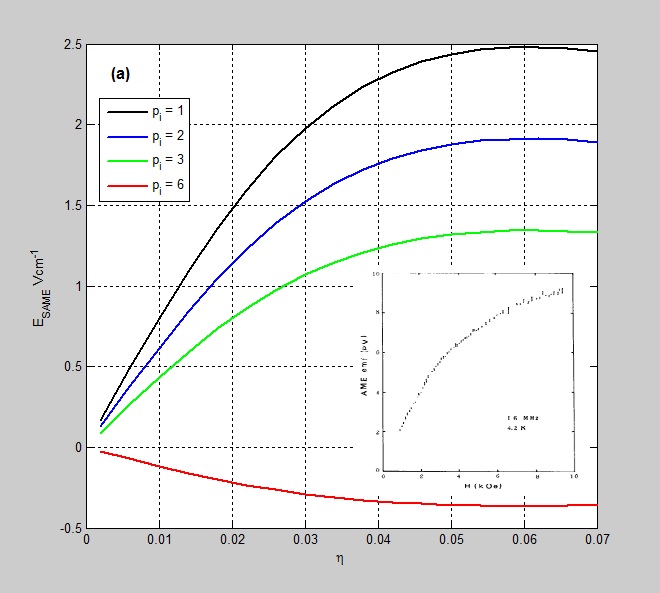}
\includegraphics[width = 7cm]{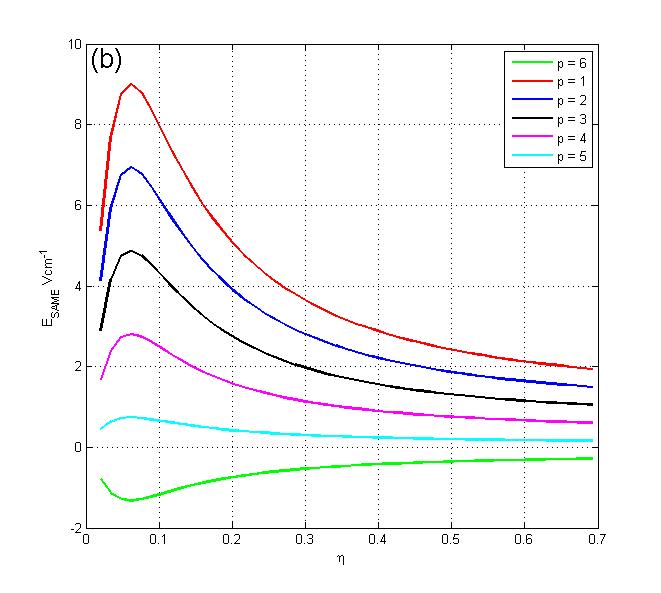}
 \caption{ Dependence of $\vec{E}_{SAME}$ versus the magnetic field strength $\eta$ for (a) $N = 7$-GNR at different sub-bands. The insert 
shows the experimental observation of $\vec{E}_{AME}$ in graphite~\cite{16}.
(b) an extended graph of $\vec{E}_{SAME}$ against $\eta$ }
\end{figure}
\begin{figure}
\begin{centering}
\includegraphics[width = 9cm]{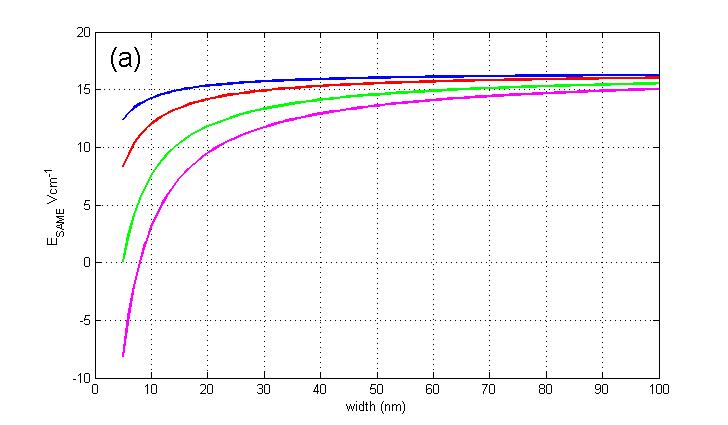} 
 \caption{(a) The $\vec{E}_{SAME}$ versus width for $p = 1, 3, 5$.} 
\end{centering}
\end{figure}
\begin{figure}
\includegraphics[width = 7cm]{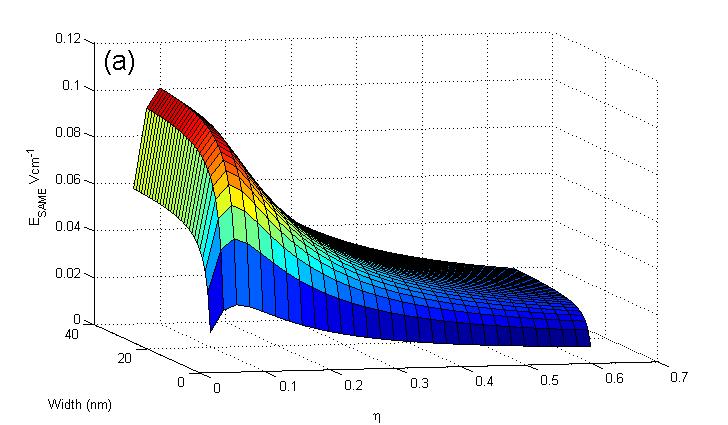}
\includegraphics[width = 7cm]{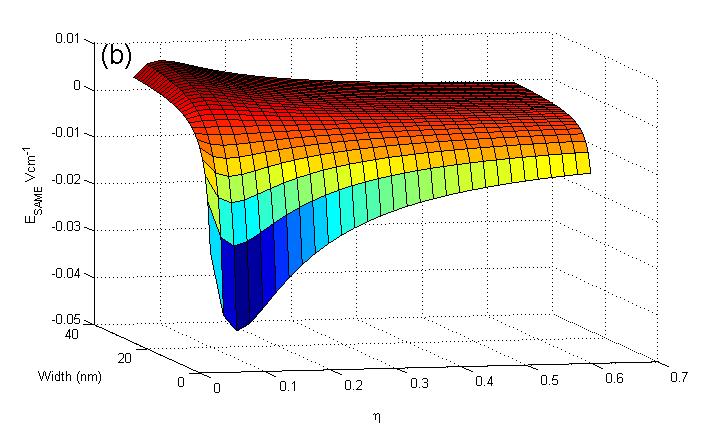}
\caption{A $3$D graph of $\vec{E}_{SAME}$ on width of GNR and $\eta$ (a) p = $1$ and (b) p = $6$.}
\end{figure}
\pagebreak
\section*{Conclusions}
The Acoustomagnetoelectric field $E_{SAME}$  in Graphene Nanoribbon (GNR) was studied. 
The dependence of  $E_{SAME}$ on the magnetic field strength $\eta$ and the width $N$ were numerically studied.
The $E_{SAME}$ obtained for low magnetic field strength in GNR qualitatively agreed with 
experimentally observed graph in graphite but for strong magnetic fields, the $E_{SAME}$ rapidly falls to a 
minimum. The graph is modulated by varying the sub-band index $p_i$ with an inversion occuring at $p_i = 6$. or the width $N$ of GNR.
At the maximum point, a magnetic field of $H = 3.2 Am^{-1}$ was calculated which is far lower than 
that measured in graphite. The $E_{SAME}$ also varies when plotted against the Width of GNR at various sub-band indices $p_i$.

\renewcommand\refname{Bibliography}

\end{document}